# Spatially Organised Membrane Flows in Biological Cells IV: Cytoskeletal Pattern Formation In Plant Cells

*How spatially-organised membrane and cytoplasmic flows establish Gravity-sensitive Microtubule/Microfibril parallelism and maintain it during growth and tropic motion*

Dr. J.T. Lofthouse A.K.C.

*MIRth Encryption Software plc, 21 Lytton Road, Springbourne, Dorset BH1 4SH. United Kingdom*

EMail: *DrLofthouse@lycos.co.uk*

## Abstract

Rather than being 'random' viscous fluids, it has been suggested that red cell cytomechanics demonstrate the flows of membrane lipids and the cytoplasm are spatio-temporally organised by convective and shear-driven mechanisms when cells are metabolically active [1-4]. Feedback between the cortical protein and phospholipid was shown to permit membrane flows to dictate the arrangement of cytoskeletal proteins in animal cells, and hence to act as the *primary* determinants of their shape and shape transitions. In this paper, by extending the new membrane model to plant cells, I show that the assumption of similar spatiotemporal organisation of the membrane and cytoplasmic fluids offers not only a feasible pattern forming mechanism for establishing the distinctive parallelism between intracellular Microtubules and extracellular Cellulose fibres, but also provides a driving force that can explain the simultaneous re-alignment of these two sets of fibres that is observed during growth, geo- and photo-tropisms.

The new model represents a 'Gestalt shift' in approaches to Cell Mechanics, as it suggests the *primary* events in biological cell shape changes are bifurcations in viscous fluid flows, rather than alterations to protein-protein interactions. It indicates that plant cell Morphogens alter cell shape primarily by altering the kinematic properties of the lipid membrane, rather than exclusively by inducing shifts in protein conformation, and explains geo- and photo-tropic shape responses in terms of perturbations to the cells' fluid dynamics. The model offers a rationale for the 'Homeoviscous Adaptation' of membrane and cytoplasmic viscosity in response to environmental stresses, and for the extensive lipid polymorphism that exists within and between species.

## Introduction

The most striking feature of plant cell architecture is the exact parallelism between their extracellular Cellulose fibres and intracellular Microtubules. Both sets of fibres lie in immediate contact with the cell membrane. Whilst many species have several layers of Cellulose fibres on their outer surface, successive layers of fibres are always found to be rotated by specific angles relative to those above them, and the innermost layer of fibres always runs parallel to and exactly mid-way between the intracellular Microtubules on the inner side of the membrane (Fig. 1). During cytoskeletal assembly and growth, Microtubules undergo a series of distinctive transverse/longitudinal re-alignments relative to the cell axis, their new orientation at each stage predicting that of the next layer of Cellulose to be deposited ([5,] see [8] for a review). Similar fibre re-orientations are also effected during growth, but the mechanism through which parallelism is established and maintained – the so-called 'Microtubule-Microfibril Syndrome' - remains unknown, and forms perhaps the most intriguing question in Botany.

It is not possible to observe cell wall deposition in 'real time' using light microscopy, but freeze etch data has yielded some indication of the sequence of events that takes place. During cytoskeletal production, protein components are produced in a given sequence. In most cases, short Microtubules appear to be the first proteins to arrive at the inner membrane surface. These line up into continuous bands of constant spacing around the cell (typically circa 28 nm - about twice the depth of the membrane in *Micrasterius*: [6,8]), and are then linked together by transmembrane MAP proteins. To date, no mechanism for producing this banding pattern has been posited. Lipid vesicles containing proteins of Cellulose-synthesizing 'Terminal Complexes' (TCs: a Cellulose synthase enzyme [9,10,11], Annexin protein (which binds phospholipids) and an ATPase in higher plants) are then produced, and fuse with the membrane surface. The consensus view is that in most plant cells, assembled TCs then move *across* the cell membrane, extruding short Cellulose microfibrils onto the external membrane surface behind them as they progress. (bacterial *Acetobacter xylinum* being an exception, since here the TC is fixed into the lipopolysaccharide layer and is immobile: [13]).

In etched replicas, Cellulose fibres are always found to lie ***mid-way between*** the inner Microtubules. TC complexes show species variations in morphology, but their spatial arrangements in the membrane can be perturbed by the methodologies used to prepare samples for electron microscopy. Each extrudes variable numbers of short microfibrils which interact laterally with each other: the distal end of each resultant fibre sticks to the Cellulose fibril above it, but the TC continues to move across the cell surface, wrapping bands of Cellulose around the cell surface like a ball of yarn. Between the deposition of each layer of Cellulose fibres, internal microtubules re-align by angles up to 90 º relative to the cell axis, ensuring that fibres subsequently deposited are rotated relative to those above them. Geo- and Photo-tropic responses (which are directional growth processes) involve similar re-orientation of both sets of fibres.

As with existing models of animal cell shape, the protein cytoskeleton is assumed to be the exclusive determinant of plant cell morphology. Cell membranes, viewed as two 'Bilayer' leaflets of 'randomised' viscous fluids, are considered to play a passive role in the process, merely adapting to the contours of the cytoskeletal structures embedded in them like soap films Since these Statics-based models liken the cytoskeleton to a proteinaceous 'Meccano' set, they assert that *changes* in cell morphology can only be driven by forces generated through changes in protein conformation [14, 15, 16]. For this reason, every mechanism of Microtubule/Microfibril co-alignment posited to date has been mechanical, assuming that proteins form 'ratchet and pinion' systems, and that the forces that generate cell shape change are generated exclusively by changes in protein conformations. These models (reviewed in [8,13]) can be sub-divided into two categories:

**A: 'Direct' hypothetical mechanisms –** these postulate that proteins form physical cross-links between the cortical Microtubules and the transmembrane proteins of the TCs. TCs are suggested to be physically 'pulled' across the cell surface by cycles of physical attachment and detachment of these links from the Microtubules. Microtubules then, are seen to act as 'guide rails' for TC motion, yet these models provide no means of establishing their distinctive parallelism and spacing in the first place. Such hypotheses are falsified by the observation that parallel Cellulose microfibrils with a stable orientation and spacing continue to be laid down on the outer cell surface even if the cells' Microtubules are first dis-assembled with Taxol [7]. The hypothesis of Seagull and Heath [162] (1980) suggested that TC units were physically linked to Microtubules, but were moved through interaction with 'contractile' Actin filaments: this suggestion is falsified by the fact that neither Cytochalasin B nor Phalloidin (which disrupt Actin filaments) disrupt oriented Cellulose deposition on the outer cell surface [17,18,19].

**B: 'Indirect ' hypothetical mechanisms -**these models do not rely on a linker protein between Microtubules and Cellulose-synthesising units, but suggest that TCs are carried across the cell in a directed 'flow' of lipid. Two mechanisms have been mooted to generate these lipid 'streams', but neither is fluid-based. Both assert the membrane remains a randomised 'BiLayer' throughout - *neither* suggests that membrane lipid flows are spatio-temporally organised in three dimensions Flows in these models are generated mechanically via the agency of a 'contractile' protein system. Hepler and Palevitz, [20] for example, suggested that a cyclical, contractile sliding of Microtubules against the surface of the membrane pushes against it, rather like an oar in water, and generates a 'flow' that transports TCs across the cell. This model provides no mechanism for establishing Microtubules in equally-spaced bands, nor of co-ordinating 'contractions' along the lengths of Microtubules, and appears falsified by two observations:

i. Treating cells with Taxol (which dissembles Microtubules) had no effect on TC motion across their surface in parallel bands [21] and
ii. In some cells (eg moss peristomes) TCs move across the cell, depositing Cellulose in straight, parallel lines without Microtubules even being present [22].

Staehelin and Giddings [23] adapted another 'Indirect' hypothesis [24], originally formulated to account for Chitin deposition on the extracellular surface of the fungi *Poterioochromonas*, for Cellulose deposition. They suggested that forces resulting from the polymerization and crystallization of the Cellulose filaments alone were sufficient to propel TCs in straight lines in the plane of the plasma membrane. In their model, extruded Cellulose fibres physically 'punt' against the surface of the membrane as it emerges from the complex, intracellular Microtubules acting as a 'picket fence', limiting the lateral motion of synthesizing units as they progress across the surface, without any protein linkage to them Whilst it is clear that emerging fibrils can produce substantial force *against a substrate* (sufficient to move a bacterial cell: [159]), this hypothetical mechanism fails to explain either how the spatial pattern of inner wall Microtubules is established to begin with, or how it is maintained during cell wall deposition.

In this paper, I suggest that the new model of the membranes and cytoplasm of metabolically active cells as fluids that are spatially organised by convective and shear-driven instabilities [1-4] offers a superior mechanism for both cytoskeletal pattern formation *and* dynamic fibre re-orientation. The 'Bilayer' structure of *animal* cell membranes is topologically falsified by the phenomenon of aminophospholipid 'Flip-Flop', which causes self-intersections. As evidence of their spatiotemporal organisation, I have suggested that observed phospholipid motions in animal cells is caused by their convective organisation when the cells' gross metabolic activity holds it in a 'Near Equilibrium' state, rather than by mechanical 'protein pumps' as suggested by others (discussed in 1,2). Phospholipid motions in the plant cell membrane are however, not as accessible to investigation because of the cell wall: consequently data regarding phospholipid 'Flip Flop' in unperturbed plant cell membranes does not exist. We therefore perform a 'Thought Experiment'. Since the new membrane model provides convincing proof of convective organisation in animal cells, we *assume* this is also the case in plant cells, and examine the effects such spatiotemporal organisation could have. By virtue of low-affinity interactions with Microtubules and Cellulose fibres, what emerges is a feedback mechanism that permits the membrane surface to act as an 'informing geometry' for structural proteins, determining their spatial arrangements, and hence the shape and size of the cell. Shear-driven bifurcations in flow planiform, induced by mechanically-coupled and spatially organised cytoplasmic flows, are shown competent to 'drag' these proteins into new arrangements, thereby inducing affine deformations of the cytoskeletal lattice and altering cell shape. This protein/lipid feedback model constitutes an 'Indirect' pattern forming mechanism by the above criteria, since it requires neither a cortical protein to act as a 'guide rail' for TC motion, nor the participation of any protein 'winch and pulley' systems, merely that the cell continuously dissipates heat internally, such that mechanically-coupled membrane and cytoplasm flows can become spatially-organised by convective and/or shear-driven mechanisms.

# Pattern Formation: How Spatially-Organised Lipid Flows Convert The Membrane Into An 'Informing Geometry' For Cytoskeletal Protein Assembly

Spatial organisation of flows in viscous fluids with strong similarities to membrane lipids can arise by two mechanisms (detailed in 24,25). In the first, gross metabolic activity produces a sufficient thermal gradient across the cell to induce surface tension-driven (Benard-Marangoni) convection. Here several flow patterns ('planiforms') – rolls, rectangles of varying proportions or hexagons– are possible. In the second mechanism, mechanical coupling between two immiscible viscous fluid layers allows one to exert shear on the other, and induce patterned flows. I suggest such a shear-driven mechanism could enable a flowing cytoplasm to exert shear on the membrane, and either produce Taylor roll flows, or if membrane flows are convectively organised to begin with, could cause flow bifurcations into roll or square cell planiforms. We assert therefore that the heat generated by gross metabolic activity is sufficient to cause phospholipid molecules in the plant cell membrane to enter a coherent state – initially a roll planiform, and that individual lipid molecules therefore traject as advancing spirals along the resultant counter-rotating rolls, with a wavelength (defined as the width of two counter-rotating rolls) approximately equal to twice the depth of the fluid layer.

We assume these rolls initially lie transverse to the plant cell axis ( Fig. 2A). Because Microtubules are acylated, and undergo low-affinity interactions with phospholipids 26 27, 28, those delivered by cytoplasmic flows to any point on the inner membrane surface will automatically be carried into the phospholipid flow 'sinks'. Once here, they will form continuous bands, laterally spaced at a distance determined by phospholipid flow $\lambda$ , and will maintain this orientation and spacing at a constant rate of internal heat dissipation. (Fig. 2B). Microtubule-binding proteins then arrive at the membrane surface. These too will be carried into flow sinks, and by virtue of acyl chains added to them post-translationally, will be **pushed** by the advancing spirals of phospholipid along the side of the Microtubules. When they encounter their corresponding association sites on two short tubules, they will bind these together into continuous transverse bands (Fig. 2C).

Lipid vesicles containing the Cellulose-synthesizing TC proteins then arrive at the membrane surface, fuse with it, and insert their transmembrane protein contents into it. These will also become trapped between flow sinks. Fluid flows within each roll are unidirectional: they will thus exert a uni-directional force (see Appendix I), and because TCs do not bind directly to Tubulin, they will be pushed across the surface of the cell as they extrude fibrils onto the outer cell surface. Cellulose fibrils also bind phospholipids with low affinity : consequently, irrespective of the position in which they lands on the cell surface, the trailing end of each Cellulose fibril will be pushed into flow *sources* on the outer cell surface (parallel to and exactly mid-way between the inner cortical Microtubules), where they can interact laterally to form the final Cellulose fibre (Fig. 2D).

This feedback mechanism is superior to previous 'Indirect' hypotheses, since it allows Cellulose to be deposited in straight, equally-spaced parallel lines on the cell surface even if Microtubules are not present. Provided phospholipid roll planiform is maintained, which requires continuousATP/GTP hydrolysis, TCs will continue to move across the cell at a constant spacing: the presence of Microtubules is not essential to the process – explaining the observations made in moss peristomes 22. This mechanism also explains the disparate effects of Taxol and Colchicine on Cellulose synthesis. *In Vivo*, both reagents cause inner Microtubule bands to disassemble, yet only Colchicine affects the deposition and parallel motion of TCs across the cell surface 21. Taxol binds to Microtubules alone, whereas Colchicine binds Microtubules *and* to phospholipid headgroups 30, and will consequently perturb their spatial organisation by surface-tension driven convection 1,2,3,4. That Colchicine perturbs cell shape primarily by its action on phospholipid, rather than by binding Microtubules is supported by its observed effect on red cell shape 1: here it causes a shape change from a biconcave disc to a sphere, despite the absence of Microtubules or contractile cytoskeletal elements. The feedback model also explains why *linear* arrays of TCs are disrupted when cells are treated with dyes that H-bond glucan polymers and phospholipids, but resume their native configuration when these are washed out - the dye causes perturbations to phospholipid dynamics and their ability to organise spatially into rolls.

# A Dynamic Growth Mechanism: Phospholipid Flow Bifurcations As The *Primary* Cause Of Microtubule/Microfibril Re-Orientation

The mechanism described above provides a motile force for TCs across the cell surface *and* permits deposition of one layer of Cellulose fibres at constant spacing in the same orientation as the inner Microtubules. In the steady state, sets of transverse bands around the inner and outer surfaces of the plant cell membrane allow them to resist lateral expansion. However, during the deposition of multi-lammelate Cellulose walls, and in response to various stimuli (gravity, red/blue light, and various plant hormones) cortical Microtubules are known to effect a series of re-orientations relative to the cell axis, accompanied by corresponding transverse/longitudinal shifts in the deposition of subsequent Cellulose fibres. How might this be achieved?

Empirically, we know Microtubule/Microfibre re-orientation is induced by diverse agents, including:-

1. The application of transient electric fields to growing plant cells [31,32].
2. Increased intracellular $Ca^{2+}$ concentrations secondary to treatment with herbicides including amiprophosmethyl, trifluraline and oryzaline [33]. Interestingly, the latter eliminates the response to gravity of amyloplasts in the tip cell of *Physcomitrella* [2034].
3. Compression or stretching (ie shear) is also a cue for re-alignment [35, 36].
4. Treatment with Auxins (eg IAA: [37], UV light [38], Kinetin [39], Abscisic Acid [40], Ethylene (re-orients from transverse to longitudinal: [41] and Gibberelic Acid (re-orients from longitudinal to transverse: [42]).

Clearly, to explain the phenomenon, we are seeking a mechanism that is sensitive to all of these perturbations. Prevailing models assume that cell shape change can only be invoked mechanically via conformational changes in specific proteins (eg Actin, Microtubules), causing extension/contraction or relative motion between pairs of structural members. However, these 'Tensegrity' models are falsified by the shape changes of the human red cell. Pertubations 1-3 listed above also induce *defined* shape changes in mature human erythrocyte [1-4] that cannot be explained by a protein-based mechanism, since these cells posses no Microtubules, and have a cytoskeleton with no 'contractile' capacity whatsoever. Using the new membrane model, I have presented evidence that red cell membrane aminophospholipid flows are convectively organised, and demonstrated how low affinity interactions with their cortical cytoskeletal components permits **bifurcations** – changes in the pattern of phospholipid flows induced by the action of these perturbations, to provide the driving force for shape changes by 'dragging' cytoskeletal components passively into new spatial configurations. In this model, the *primary* effect of all four perturbations listed above is on lipid flow dynamics. As detailed below, the feedback mechanism suggests there are essentially three ways in which phospholipid flow patterns (and hence plant cell fibre orientation) could be altered in the membrane of a metabolically active cell.

1. **Shear-Induced Microtubule/Microfibril Re-Orientations**

    A classic bifurcation sequence in the annals of Fluid Dynamics is a pitchfork 'hexagon to roll bifurcation' that arises when mechanical shear is applied to the surface of a convectively organised viscous fluid ([43]:**Fig. 3A**). A pattern of hexagonal convection cells bifurcates first into a series of rolls running *transverse* to the direction of applied shear, then, as shear rate increases, flows pass through a square cell intermediate and eventually form counter-rotating rolls with the same wavelength running *parallel* to the applied shear. With the inner MT and outer Cellulose fibres effectively 'slaved' to phospholipid flow sinks, **Fig. 3B** demonstrates in a planar case how the application of mechanical shear would cause a reorientation of phospholipid rolls that would drag both inner and outer fibres into a new alignment, whilst maintaining their spacing.

    In three dimensions, if we assume that membrane phospholipid is convectively-organised, and that phospholipid molecules traject on the surfaces of rolls running transverse to the cell axis with a wavelength $\lambda$, bands of Microtubules, pushed into flow sinks will adopt a spacing of $\lambda$. Assuming the membrane and cytoplasm are mechanically coupled, the latter can exert shear force on the membrane, inducing a bifurcation in phospholipid roll orientation by 90 º. Microtubules, caught in flow sinks, will re-orient with them, breaking connections with transmembrane proteins, form new connections when parallel to the cell axis, and defining the orientation of the next layer of Cellulose fibres on the outer surface.

    Plant 'Morphogens' are known to have an effect on both the direction and velocity of cytoplasmic streaming (eg [44]), as do anaesthetics [149]. Under the new model, this appears therefore to be an *indirect* effect, reflecting the fact that these hormones alter membrane lipid viscosity, which changes the mechanical coupling between the membrane and cytoplasm, and in effect, reduces the 'brake' on cytoplasmic flow velocity. This would alter the shear exerted on the membrane. At low shear, phospholipid vortices run transverse to the cell axis: at

higher shear, these re-orient, carrying cytoskeletal elements with then into a new orientation. Sequentially increasing and decreasing mechanical coupling enables sequential layers of cellulose to be deposited with alignments rotated by an angle up to 90 degrees.

2. **Protein Conformational Shift Dependent Microtubule/Microfibril Re-Orientations**

Cytoskeletal proteins form 'corrugated' surfaces past which the lipid and cytoplasm flow: protein-lipid feedback thus allows changes in protein conformation, and post-translational protein modifications, to feed into cellular fluid dynamics and to contribute to **cumulative** instabilities in phospholipid planiform , but in a manner that is not limited to a Statics-based 'lock and key' mechanism in which conformational shifts alter their ability to bind to other proteins.

The *phosphorylation* of protein residues effectively adds negative charges to them at positions determined by sequence. Because phospholipid headgroups are also negatively-charged, newly added phosphate groups on the proteins that they run across will cause deflections in their stable trajectories, generating instability or 'wobble'. These effects would be cumulative, and above a certain threshold, could induce a complete bifurcation in the planiform of phospholipid flow through the membrane above them. This would alter the spatial distribution of cytoskeletal proteins by feedback, and hence effect a change in cell shape. Most of the enzymes that catalyse the addition of phosphate groups to proteins are themselves ATP or GTPases – that is , they release heat into the cells' fluids as they add these groups. This localised heating at specific points on membrane-bound proteins will also effectively 'force' lipid convection at points in the membrane above them, which will also deflect phospholipid flows from a stable convective or shear-driven pattern, and change cell shape: I suggest this re-reading is actually how secondary messenger molecules alter cell shape. The extent of β-Tubulin phosphorylation is known to vary during the plant Cell Cycle and differentiation 46, 47, and a known plant cell morphogen, UV light, also induces phosphorylation of membrane-associated proteins *In Situ*48.

The post-translational ***Deamidation*** of Asparagine and Glutamine residues also changes the contours of proteins that run against the membrane. Loss of amide groups converts these amino acids respectively into negatively-charged Aspartic- and Glutamic Acid. It produces a 180º 'kink' in the protein backbone, and also changes the charge distribution along the polypeptide. It is universally observed in most proteins of all cells throughout their entire Developmental trajectory 167, has been reported in Tubulin 50, and could therefore contribute to the observed age-dependent shifts in Microtubule orientation 51,52, though senescence is also accompanied by changes in the kinematic viscosity of membrane lipid (through phospholipid oxidation or remodelling) which could also generate 'wobble' in stable phospholipid trajectories through the membrane 53.

3. **Plant Hormone-Induced Microtubule/Microfibril Re-Orientations**

Plant cell 'Morphogens' such as Ca $^{2+}$ and hormones are currently assumed to alter cell shape exclusively by binding to and changing the conformation of cytoskeletal proteins (eg 'Tensegrity' models:14,15,16). The new feedback mechanism suggests this is not the case: all have long been known to affect the viscosity of the lipid membrane, but these effects have previously been marginalised. Under the new model, these effects can be seen to alter the degree of mechanical coupling between membrane and cytoplasm, the magnitude and direction of shear application to the membrane, and hence to act as the driving force for the cytoskeletal re-arrangements that cause shape changes. The effects of plant cell morphogens on lipids include:-

- **Ca $^{2+}$** - this activates an ATP-dependent Ca $^{2+}$-pump, which itself hydrolyses ATP as it expels cations, adds to the increased rate of internal heat dissipation, which increases the temperature gradient across the cell. Ca $^{2+}$ also binds phospholipid headgroups, destroying their ability to form the 'proton transfer complexes' through which they are held in a coherent convective state (see 1, 2). On prolonged exposure, Ca $^{2+}$/ionophore also stimulates the production of 'Reactive Oxygen Species in the membrane (ROS:54). These react with unsaturated lipids and produce malonyldialdehyde (MDA: 55, which cross-links the amino groups in both protein and PS headgroups together 56, and consequently increases membrane viscosity. This is known to perturb the rate of aminophospholipid 'flip flop' through the red cell membrane. In the red cell, because PS and PE fractions are richer in unsaturated fatty acids than either PC or SM, they are more prone to such oxidative damage. Because convective onset is a function of viscosity, such cross-linking would also have a dramatic effect on the planiform or wavelength of lipid flows. The feedback model presented in 3 explains why red cells treated with MDA become undeformable, and refractory to shape change57: aminophospholipids cannot flow, and hence can no longer re-position the cytoskeletal proteins .
- **Herbicides** such as trifluraline and oryzaline 33 increase intracellular Ca $^{2+}$ levels.
- **Abscisic acid** induces a Ca $^{2+}$ decrease in protoplast cytoplasm 40, reversing the above effects. It also reverses the effects of Gibberellic Acid treatment on epicotyls, eliminating the induced predominance of transverse Microtubules.
- **UV light** exposure induces lipid peroxidative damage 58,59. This induces the production of MDA, which cross-links aminophospholipid headgroups together, and to protein, and would hence alter the effective viscosity of the membrane, impeding phospholipid flow and spatial organisation.
- **Auxins** bind phospholipids 30, and *In Vivo*, *reduce* MDA levels in cell membranes and hence act to reduce membrane viscosity 60. Different species of plants react to different amounts of Auxins, a fact used to advantage as a method of weed control, that I suggest is as attributable to differences in the lipid composition (and hence viscosity) of their membranes as it is to differences in their cytoskeletal protein components.
- **Kinetin** protects against oxidative and glycoxidative damage generated *In Vitro* by sugars and by an iron/ascorbate system, inhibiting both the formation of BSA-carbonyls after oxidative damage and the advanced glycation end product 61, indicating it is part of a feedback loop that maintains normal membrane viscosity.
- **Ethylene** applied to epicotyls and hypocotyls can induce a transverse to longitudinal change in Microtubule/microfibril orientation 45,41,62
- **Mechanical stimuli** (wind, touch) stimulate the production of Ethylene, which promotes the same effect when added exogenously. It stimulates the expression of a lipase (lipolytic acyl hydrolase) normally expressed at the onset of petal senescence in Dianthus caryophyllus. This de-esterifies fatty acids in both *In Vitro* and *In Situ* assays of enzyme activity, and since the viscosity of lipid is primarily a function of acyl chain composition, may also affect the angle which phospholipid roll sources make relative to the cell axis. De-esterification of membrane lipids and ensuing loss of membrane structural integrity are well established early events of plant senescence, and the expression pattern of this lipase gene together with the lipolytic activity of its cognate protein indicate that it plays a fundamentally central role in mediating the onset of senescence 63.
- **Gibberellic acid** change the orientation of cortical Microtubule from longitudinal to transverse in maize 42, onion leaf-sheath cells and pea seedlings 64. These effects can be reversed by Colchicine 65.
- **Colchicine** treatment of elongating semicells from unicellular green alga *Closterium* reportedly leads to loss of Microtubules from the wall, and deposition of Cellulose fibres in 'random' orientations: cells lose their normal tapering cylinder morphology and become spherical 79. This reagent destabilises Microtubules *In Vivo*. *In Vitro*, however, it only binds non-covalently to *unpolymerised* Tubulin dimers, not to Microtubules 66,67. The assertion that this substance affects Microtubule/fibril orientation, and thus plant cell shape by direct association with Tubulin is falsified by it's ability to invoke 'sphering' of human red cells 68, which retain no Tubulin or contractile elements whatsoever. Colchicine binds phospholipid 69, strongly suggesting that it's action in plant cells is primarily to effect alterations in phospholipid dynamic, and to perturb their spatially-organised flow patterns.

Colchicine disturbs the ability of electric fields 70, Auxin 37 and Gibberellic Acid 71,72 to induce Microtubule/Microfibril re-orientations, suggesting all are operating on the same part of the cell, feeding into the fluid dynamics of the cell. The opposing effects of Absisic Acid and Gibberellic Acid on the orientation of wall fibres is explained by the fact that one increases mechanical coupling between the cytoplasm and membrane, the other decreases it.

# A Mechanism For Gravitropic Bending

It is clear that gravitationally-induced instability is a source of biological information 73,137, but it is difficult to see how such a small force could play a role in biological cell systems – the molecular processes by which they perceive and transduce gravity are poorly, if at all, understood. Whilst gravity is experienced continuously by *all* biological cell systems, and all cell types (animal, plant and bacteria) change shape on exposure to microgravity, gravitropic plant cell growth processes are the most widely studied. The two most obvious effects of reducing the gravity vector are (i.) alterations to the velocity and direction of cytoplasmic streaming, and (ii) alterations in the direction of plant cell growth. When re-interpreted against feedback mechanism of cell shape, extant data give a strong inference that fluid mechanisms act as the primary cause of gravi-sensitivity in cell shape and dynamics. These, I suggest are due to buoyancy-driven cytoplasmic flows. Many other aspects of plant growth and Development are affected in a micro-gravitational environment, these include inhibited growth, reduction in cell division, and dis-oriented growth. Current models *assume* that the growth hormone Auxin is involved, however, it has recently been shown that corn plants grown in microgravity do not exhibit any overall changes in Auxin levels 74,154 , giving credence to the model presented in this series of papers, whereby a reduction in the gravity vector alone can lead primarily to either a change in the pattern of buoyancy-driven convection of the cytoplasm, altering the area(s) of the membrane subjected to shear, and determining the orientation of cytoskeletal elements, and the shape and future growth direction of the cell.

The simplistic physical assumption on which *all* existing models of plant gravitropism are based is that 'gravity acts downwards' (!), however tropic responses to gravity can be positive, negative, transversal (90 degrees to the vector) or plagiotropic (non-perpendicular to the vector). As with cytomechanical models of animal cells, the plant cell membrane is also assumed to act as a passive 'rubber sheet', and cell motion is attributed exclusively to the activity of its' 'contractile' protein cytoskeleton. The most widely accepted model of gravitropism, the 'Starch-Statolith Hypothesis' (reviewed in 75, suggests that gravity acts on solid 'objects' that are attached to the cytoskeleton. As these fall under the influence of gravity, they are assumed to pull on specific trans-cytoplasmic protein filaments, and exert a force. Under this scenario, these objects "must be of sufficient mass to ensure the work done on them is higher than the elemental kinetic energy to discriminate between gravity-dependent movement and random thermal background " (Sievers et al., 1991). Problems are evident when one closely examines these data, however, as the 'perception times' for gravity response in Lepidium roots 75, 76, and in Avena coleoptiles 77 are significantly lower than even the threshold time necessary to distinguish gravity effects from 'thermal noise'. According to published and widely accepted calculations 78, of all plant cell inclusions, only the starch-producing amyloplasts have enough 'potential energy with respect to gravity' to perform the work deemed necessary for gravity perception.

Not surprisingly, no extant model explains the gravity response of the plethora of cells which lack amyloplasts [80],[81], or the presence of amyloplasts in cells that are gravitationally-inactive (eg Guard cells/secretory cells), or the gravity-dependent polarity or velocity of cytoplasmic streaming [82],[83], or why disruption of cellular Actin (the major 'contractile' protein for any 'winch and pulley' mechanism of the change in cell shape that accompanies gravitropic bending) has no effect on the gravitropic response [130].The feedback mechanism does.

**All** gravitropic plant responses in a 1G environment involve a *change* in the direction of cell growth – this in turn demands a differential elongation rate on opposite sides of the cell. There is a clear difference in the orientation of cortical Microtubules between cells on the concave and convex sides of phototropically- and geotropically curved corn coleoptiles. Nick et al (1990) clearly demonstrate this in horizontally-oriented sunflower hypocotyls, where differential growth correlates with the re-orientation of cortical MT/MF from transverse to longitudinal in the upper epidermis, whilst those in the lower epidermis maintain their original transverse direction. According to the new model of spatially organised cellular fluids, this demands a 'transverse to parallel' re-orientation of lipid rolls trajectories in specific areas of the membrane, and this demands that different areas of the membrane experience different shear exposure.

The feedback model suggests two possible scenarios could result in areas of the membrane being subjected to differing shear forces:-

**A:** It has been observed that when cells with a vacuole are placed horizontally or tilted, the buoyancy-induced lateral displacement of the vacuole makes the layer of cytoplasm thicker on the lower side of the cell [81]. Under my model, because cytoplasm and membrane are coupled, this displacement of the vacuole will increase the shear force exerted by streaming cytoplasm on the *upper* membrane relative to that beneath the vacuole. Phospholipid flows in the upper membrane will therefore preferentially undergo a localised bifurcation in phospholipid roll orientation from transverse to longitudinal [43]:**Fig. 3A**), permitting cell extension. Because Microtubules are joined together by transmembrane MAP proteins, those in the upper section of the cell will follow the new phospholipid flow sinks by breaking linkages, and form ligations in the new orientation. This will cause the cells to bend **downwards** as the cells grow (ie will induce positive gravitropic bending)(**Fig. 4A**).

**B:** Cytoplasmic flows obviously differ in plant cells that do not have a vacuole. As shown in **Fig. 3C,**, in 'Natural' (ie unforced) buoyancy-driven convection, the pattern of fluid flows within a sphere of viscous fluid heated from a central source is a function of viscosity, temperature, gravity vector and importantly, its radius [85, 86, 87], [88,89], [90, 91,92]. If we assume that a 'sphere 'of cytoplasm is encased by a concentric 'sphere' of mechanically-coupled membrane, the pattern of up-and down-welling cytoplasmic flows will determine which areas of the membrane experience a shear force, and the direction in which this is applied. In Natural convection, stable solutions can be axi-or non axisymmetric. A non-axisymmetric pattern of flowing cytoplasm could therefore exert shear only on one area of the membrane, invoking localised re-orientations of cortical proteins, causing the cell to 'bend' asymmetrically, upwards, downwards, or at any angle between.

Flow patterns manifest in a biological cell will obviously be much more complicated that the Spherical Benard case. Cells are rarely spherical, and because we are not actually dealing with a 'Natural' convection situation, both membrane and cytoplasmic fluids will develop vorticity as they flow over the corrugated surfaces of cytoskeletal proteins. The points of heat injection into the fluid 'sphere' will never be exclusively centralised as in the theoretical treatments of the Spherical Benard Problem, since ATP- and GTP-ase proteins bind to the cytoskeletal meshwork at specific points in space. In a real cell also, membrane viscosity is a function of lipid composition: this varies from cell to cell, and is adjusted continuously throughout lifespan. Cytoplasmic viscosity too varies throughout lifespan and in response to environmental stresses. All impossible to model quantitatively - however, the point to be made here is that non-axisymmetric cytoplasmic flows are observed, and are realisable by this mechanism. Suffice to say that a non-axisymmetric upwelling of cytoplasmic fluid will generate shear at a specific region on the membrane surface. Phospholipid roll trajectories through the membrane in contact with these areas will, under the feedback mechanism, preferentially re-orient, and drag Microtubule/Microfibrils into co-alignment relative to the cell axis. The direction in which the cell 'bends' - upward, downward, or at any angle between - is determined by cytoplasmic flows. As the pattern of cytoplasmic flows is determined by factors such as the kinematic viscosity of its' membrane (lipid composition) and the viscosity of its' cytoplasm, Gravitropic response could change during lifespan: the Dynamic Template mechanism is therefore superior to preceding models of Gravitropism because...

- It allows a positive, negative, or plagiotropic cell shape response, since the region of membrane that experiences shear is a function of the cytoplasmic flow pattern.

- It offers an explanation for the observed age-dependent shifts in Microtubule orientation [51], [52], and the decreased gravitational response of aged plant cells. This can be attributed to either changes in the kinematic viscosity of membrane lipid (through oxidation or remodelling) and/or age-related deamidation of residues in Tubulin *In Vivo* [50] causing 'wobble' in stable phospholipid trajectories through the membrane, as suggested in the red cell membrane. As suggested in preceding sections, the addition of negatively-charged phosphate groups to proteins that lay against the membrane will cause deviations in the trajectories of PS/CL headgroups moving past them, hence generating 'wobble', whilst peroxidative damage induces the production of MDA, which cross-links aminophospholipid to protein, and would hence affect the viscosity of the membrane,the extent of mechanical coupling between cytoplasm and membrane, and

- would therefore affect cytoplasmic streaming rate and potentially its' direction as well.

- It explains the inhibition of plant cell gravitropic responses at 4 ºC 93, and for the perturbation of this response caused by treating cells with Colchicine. Obviously, alteration to environmental temperature affects the temperature gradient across the cytoplasm and membrane, and will affect convective flows in both. Colchicine only binds to unpolymerised Tubulin dimers 66,67, yet extracellular Cellulose MF's are observed to re-align when this reagent is added to intact plant cells. The assertion that this substance affects plant cell shape changes exclusively by direct association with Tubulin is falsified by it's ability to invoke 'sphering' of human red cells68, which retain no Tubulin or contractile elements whatsoever. Colchicine interacts with phospholipid 69, strongly suggesting its action *In Vivo* involves the inducement of bifurcations in lipid flow patterns, and that its effect on gravitropisms are due to lipid events.

- It offers a feasible explanation of the plant cell's ability to respond to gravity changing during the Cell Cycle 94. Before M-phase, Gravitropic bending not only ceases, but appears to reverse: it would be interesting to establish whether this ties in with changes to the composition of the lipid membrane during Development 95, 96. Cytoplasmic streaming rates and patterns also vary from cell to cell, and during Development 97, 98. A hypothetical model of Morphogenesis, whereby instabilities in lipid organisation act as the primary cause of Symmetry-Breaking instabilities in biological systems is delineated in 1, 2.

# A Mechanism For Phototropic Growth

The new model also offers an explanation for the interplay that is observed between Photo-and Gravi-tropic plant cell responses, by providing a common mechanism into which both processes can channel. Naturally produced Auxins such as IAA invoke longitudinal to transverse MT/MF re-orientation relative to the cell axis 99, 100,84: see 65 for a review). They bind phospholipids 30, and *In Vivo*, effectively 'mop up' the damage caused by oxidative processes - ie they reduce MDA levels in cell membranes and hence alter membrane viscosity 60; their effect on lipid peroxidation is one of the most rapid biochemical responses known. Treatment of cells with UV light leads to a predominance of microtubules and cellulose fibrils that lie *longitudinal* to the cell axis, induces phosphorylation of membrane-associated proteins *In Vivo*48, and lipid peroxidative damage 58, 59. The time scale on which it causes MT/MF re-orientation in Adiantum protonema 38 is a mere ~10 minutes 101.

Phototropic bending, under this model, arises because light destroys Auxin. Treatment of cells with Auxin usually favours longitudinal alignment of Microtubules relative to the cell axis: it bind phospholipids 30, and *In Vivo*, reduces MDA levels, and hence membrane viscosity 60; its' ameliorating effect on lipid peroxidation is one of the most rapid biochemical responses known. Contrariwise, UV light *induces* lipid peroxidative damage 58, 59, and induces the production of MDA, which cross-links aminophospholipid to protein, and would hence increases membrane lipid viscosity. At a constant rate of metabolic pumping therefore, localised changes in the orientation of cortical tubules could be brought about by localised change in the planiform of convectively organised phospholipid. The feedback mechanism readily explains positive phototropism - light destroys Auxins; the cells on the side of a plant exposed to light will not divide or grow as quickly as those on the shaded side and thus the plant will bends toward the source, as if 'growing towards it' Fig. 4C.

Supporting evidence for the feedback mechanism it found in the observation that treatment of cells with Auxins or UV light eliminates the gravity-dependent polarity of cytoplasmic streaming 82: the Dynamic Template suggests that UV light affects cytoplasmic streaming patterns *indirectly*, by exerting a primary effect on the the lipid membrane, altering the amount of mechanical coupling between it and the cytoplasm in specific areas of the membrane, altering the amount of 'damping' that the flowing cytoplasm experiences at the same time as it changes cell shape. Some plant cells, however, exhibit a *negative* phototropic response. The model suggests that paying closer attention to cellular fluid dynamics will allow the full elucidation of these biological mechanisms.

___________________________________________________________________________

# Discussion:

This paper suggests that far from being 'random' fluids, both the cytoplasm and membrane of metabolically active biological cells can, by virtue of convective and/or shear-driven organisation, play a central role in establishing the structure of the cytoskeleton, and hence the mechanical responses of the mature cell. It demonstrates how a convectively-organised cell boundary and cytoplasm provides the cell with a dynamic pattern-froming supplement to Reaction Diffusion schemes 102, and a mechanism that, through feedback, can produce continuous, gravity-dependent Symmetry-Breaking changes in cell shape throughout the entire Developmental trajectory.

# Implications for Plant Cell Development

Plant cell Development demands a mechanism that allows these re-orientations be 'timed' to occur at specific intervals, and in response to external signals as diverse as light and changes in the gravity vector. The new feedback model of cell shape allows the amount of shear exerted on the membrane to be continuously varied during Development without necessarily involving alterations to protein conformation, though these can contribute to changes in the cells' fluid dynamics via feedback. Whether or not cytoplasmic flows will exert sufficient shear to induce a bifurcation in phospholipid flow patterns in the membrane above them will depends on the magnitude of the shear force applied and the direction of application – this in turn will be determined by the pattern and velocity of cytoplasmic flow, and the extent of mechanical coupling between the two viscous fluid layers

Empirically, *thermal* coupling appears to be favoured in viscous multilayer systems when the buoyancy forces in them are of similar strength, whereas *mechanical* coupling dominates when buoyancy forces differ 103,104. The buoyancy force of a viscous fluid is a function of its' average mass density, thermal expansion coefficient, the gravity vector, and the temperature gradient across it (assuming depth is constant). Qualitatively, in a cellular context , the mass density of the membrane is determined by it's lipid composition, whilst that of the cytoplasm is determined by water content and the concentration of 'compatible solutes' such as mannitol or glycerol. The temperature gradient across the cell is a function of the amount of ATP/GTP hydrolysed per unit time, and environmental temperature, as is cytoplasmic streaming rate.

The amount of shear experienced by the membrane can therefore be varied continuously during plant cell lifespan because membrane lipid composition (and hence microviscosity) is continuously altered in the following ways:

(i) Microtubules 28 and other cortical cytoskeletal proteins bear acyl chains along their entire contour length consequently, as these incorporate into the membrane, they will insert into the membrane and alter its' viscosity.

(ii) Cellulose synthesizing TCs themselves are delivered to the membrane in lipid vesicles. These fuse with the membrane as they insert their contents across the membrane, which will also continuously alter viscosity. TCs have a lifetime of ~20 minutes, and are produced continuously during cell wall deposition 105.

(iii) More lipid is added during Development.

(iv) The lipid composition of biological cells is altered in response to environmental stresses (Homeoviscous Adaptation)

(v) The temperature gradient across a cell is a function of the amount of ATP/GTP being hydrolysed internally (metabolic rate) and environmental temperature: this is also a variable in Development. *In Vitro*

In sharp distinction from other models, feedback onto the genome is possible, since the feedback model treats the nucleoplasm/nuclear membrane/cytoplasm/plasma membrane and environment at five mechanically-coupled layers of viscous fluids, and DNA interacts directly with the nuclear membrane and cortical scaffold proteins: this offers an explanation of 'shock protein' expression in response to alterations in environmental temperature, pressure, viscosity and exposure to microgravitation. The convective organisation of cytoplasm by the Rayleigh Benard mechanism thus provides a means of transducing gravitational information into biological form because it is mechanically coupled to another immiscible layer of fluid which is also spatially organised, to which cytoskeletal protein is slaved. A hypothetical model of Morphogenesis, whereby instabilities in lipid organisation act as the *primary* cause of Symmetry-Breaking instabilities is detailed in 1,2.

In stark contrast to other models, the effect of cytoplasmic shear forces on a convectively-organised membrane provides a fluid-driven mechanism that could even reproduce the intriguing helicoidal Cellulose/Microtubule alignments observed in the unicellular green algae *Oocystis* 122 and in several other cell types such as root hairs and pollen tubes 6,97,107,108:**Fig. 6**. In Allium root hairs, for example,the deposition of crossed helical (+/- 45 º) MF appears to be controlled by a shifting helical MT, which also form 45 º helices 109, 110, but no mechanism for this shift has previously been posited. Helicoidal vortice patterns (ie spiral flows) can arise in Couette-Taylor cylinders, where they are thought to be caused by radial temperature gradients ( see 24,25for a review: also 111 - 117,160. Tominaga et al. 44have observed *helical* cytoplasmic streaming patterns in root hair cells, and that these change into different patterns when the cell is treated with plant hormones. As shown in **Fig. 6**, changes in the direction of cytoplasmic streaming will induce re-orientation of phospholipid roll patterns through the membrane will permit the second layer of Cellulose MF to be constructed underneath the first, but at an angle of 90 º. Additional shifts could produce cell walls with three rotational centres, as seen in *Glaucocystis*. There is a dearth of detailed observation regarding the *patterns* of cytoplasmic flows: the shear mechanism demonstrates that the direction of shear is as relevant as its' magnitude to its' ability to cause bifurcations in membrane phospholipid flows. Clearly research that makes more detailed observations is warranted

It is worth emphasizing that most of the visual data that has accumulated on plant cells has been obtained using Electron Microscopy. This involved freezing the cell, and/or treating the sample with chemical fixatives such as Glutaraldehyde which are known to perturb phospholipid phase behaviour. The images with which we build models of the plant cell cytoskeleton may not therefore reflect the spatial arrangements of cytoskeletal components in the 'Far from Equilibrium' state: we cannot be absolutely certain that the patterns we see are those that actually exist *In Vivo*. Glutaraldehyde cross-links phospholipid headgroups to protein, and protein to protein, and treatment will perturb stable phospholipid flow patterns through the membrane. Prior fixation with this reagent or cryoprotectants such as Glycerol may therefore affect the planiform and/or wavelength of convectively organised phospholipid, and alter the distribution of proteins that are seen in micrographs. The sensitivity of plant cell shape to changes in phospholipid dynamics may explain discrepant reports in the literature concerning whether Cellulose synthesising TCs form 'hexagonal' or 'row' arrays on the cell surface (see 118, **Fig. 6**). According to Giddings and Staehelin 8, hexagonal arrays are only seen in intact tissues which have been ultrarapidly frozen *without* prior exposure to aldehydes or cryoprotectants such as glycerol, and are generally not found in protoplast systems or cells grown in suspension cultures (Brown9: see his website at www.botany.utexas.edu/facstaff/facpages/mbrown/ ). Linear arrays have been reported in several algae 9,118,, 119, but hexagonal arrays have been recorded in others (eg *Zygnematales*) and vascular plant cells. Under the feedback mechanism, differences in the planiform of convection (rolls or hexagons) are anticipated to arise as a result of differences in the viscosity of the fluid membrane – it is conceivable that different cell types could manifest different stable planiforms when metabolically active because of subtle differences in their lipid acyl chain composition. Since differing types of algae have either hexagonal *or* row arrangements of TCs, and their membranes have different lipid compositions (acyl chains), it is possible that the stable RB/BM convective planiform in some species are hexagonal, rather than roll patterns, or hexagonal planiforms are stable during periods when the cell is not producing Cellulose. The spacing between linear rows is the same as the centre-to-centre spacing between hexagonal arrays, ~28 nm 6,8. Linear convection theory predicts convective **onset** should produce a lipid roll wavelength ≥ 2d 25: hence a protein pattern wavelength of 28 nm is in good qualitatively agreement with a membrane depth of ~10 nm. Shear induced 'hexagon to roll' bifurcations in convective fluid 43 do not change wavelength: *This does not preclude the possibility that hexagonal wavelengths greater than ~20 nm arise, only that wavelengths shorter than ~20 nm could not).*

## A rationale for 'Homeoviscous Adaptation' to environmental temperature changes

The reliance of cell shape on the kinematic viscosity of the cytoplasm and membrane provides a reason for the evolutionary survival of cells whose feedback loops allow them to maintain these parameters when subjected to environmental stressors. Whereas animal cells are mobile, higher plants cannot move to avoid adverse environmental conditions, and have to alter their Developmental mode to assume a form appropriate for habitat and climate. Plants that exist in regions of northern latitudes endure dramatic seasonal temperature changes, which can span 70-80°C. At both extremes, the effects can induce dehydration, protein inactivation and, in extreme cases, cell death. Most plant and animal cells respond to decreased environmental temperature by effectively *lowering* the phase transition temperature of their membranes and altering the viscosity of their cytoplasm. – a process known as 'Homeoviscous Adaptation' (reviewed in 120)

In the membrane, all identified mechanisms involve feedback loops that conspire to maintain 'effective' membrane viscosity by making appropriate alteration to the proportion of unsaturated acyl chain species within lipid subclasses. Identified mechanisms include:

i. redistributing acyl moieties within lipid classes
ii. cis/trans isomerisations
iii. altering the length of acyl chains
iv. alteration of the proportion of PS/PE in membranes
v. changing the lipid/protein ratio, either via alteration to protein production rate or via loss to the environment.

Cytoplasmic kinematic viscosity appears to be altered by the production of various solutes. *Pyramimonas,* for example, a pyramidal-shaped alga, is found in both temperate and polar oceanic waters, and also inhabits saline lakes on the coast of Antartica. At temperatures as low as –10 °C, this unicellular organism retains its characteristic shape to at least –14 °C.and will grow at temperatures as high as 16 °C because it is able to modify its intracelullar environment . Eukaryotic algae that live in hyperosmotic conditions use polyols as compatible solutes to balance their internal osmotic pressure with that of their surroundings. eg *Dunaliella* accumulates glycerol, and *Platymonas* accumulates mannitol . Glycerol has been shown to stabilize intact Microtubules , but the new mechanism suggests an additional reason for its' production. In the absence of a capacity to adapt, reducing the environmental temperature would have a profound effect on pattern formation in the membrane, and hence on cell shape and information processing ability. In mammals, an inability to adapt is associated with hibernation.

______________________________________________________________________

# Figures and Tables:

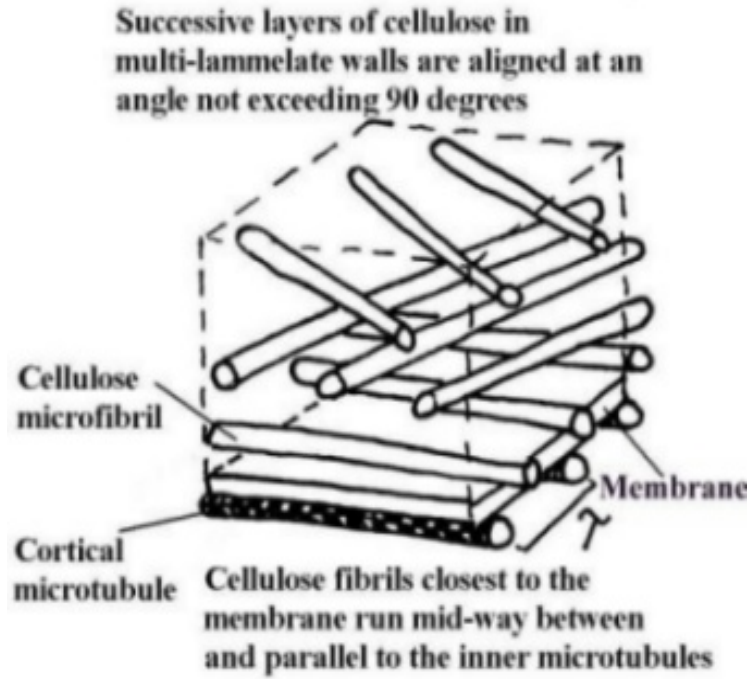

Fig. 1: Schematic representation of the plant cell wall.

Cortical Microtubules run in equally spaced bands around the inside of the cell membrane. Cellulose synthesizing units lay microfibril cables *exactly mid-way between them* on the outer cell surface, by an unknown mechanism. Each layer of external fibres is laid down at an angle not exceeding 90 º to that beneath it. The cortical Microtubules and innermost Cellulose fibres always share the same alignment, relative to the cell axis, and re-align simultaneously during growth and tropic bending. Under prevailing 'Statics-based' models, the membrane is assumed to be two concentric surfaces, each homeomorphic to a sphere (a 'BiLayer'), on which phospholipids are free to rotate and traject laterally, but with entirely 'random' trajectories (after [8]).

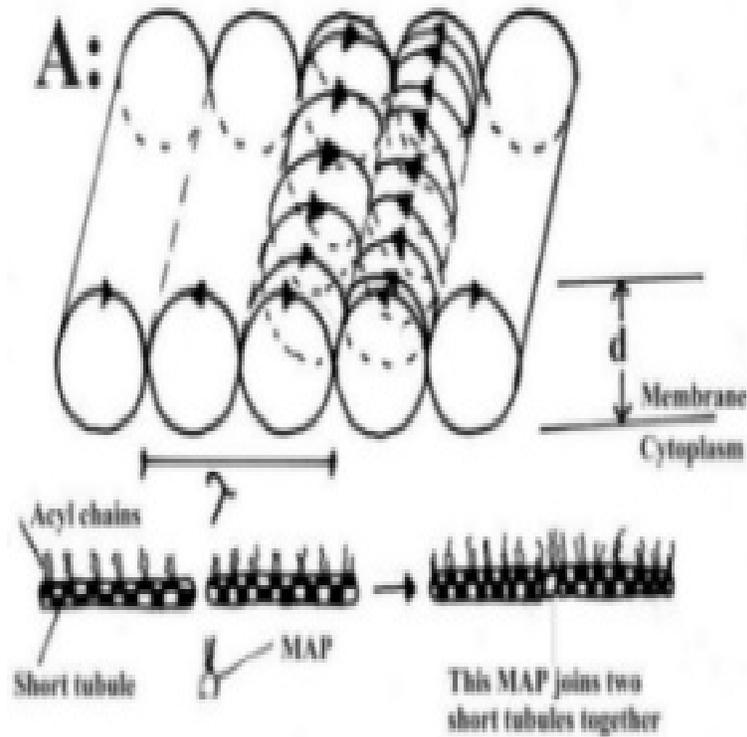

## Fig. 2: Spatially-Organised Phospholipid Flows as an 'Informing Geometry' for Cytoskeletal Assembly

**A:** The feedback model asserts that intracellular heat dissipated by gross metabolic activity is sufficient to cause spatio-temporal organisation of membrane phospholipid flows into a pattern of counter-rotating rolls, either by a surface-tension-driven convection, or by shear, exerted on it by a convectively organised cytoplasm. Flow wavelength ($\lambda$) is defined as the distance between two flow sources or sinks, and is in the region of twice the depth of the fluid layer. The trajectories of individual phospholipid molecules within shear-induced rolls are advancing spirals on the surface of individual rolls. Whilst biological membranes are usually said to have a depth of 10 nm, it must be borne in mind that depth is measured in membranes that have been fixed and frozen, and are therefore no longer in the 'Near to Equilibrium' state of the metabolically active cell. Membrane depths quoted in the literature vary from circa 4 – 19 nm in different cell types, and are strongly influenced by the amount of water present and fixation methods used.

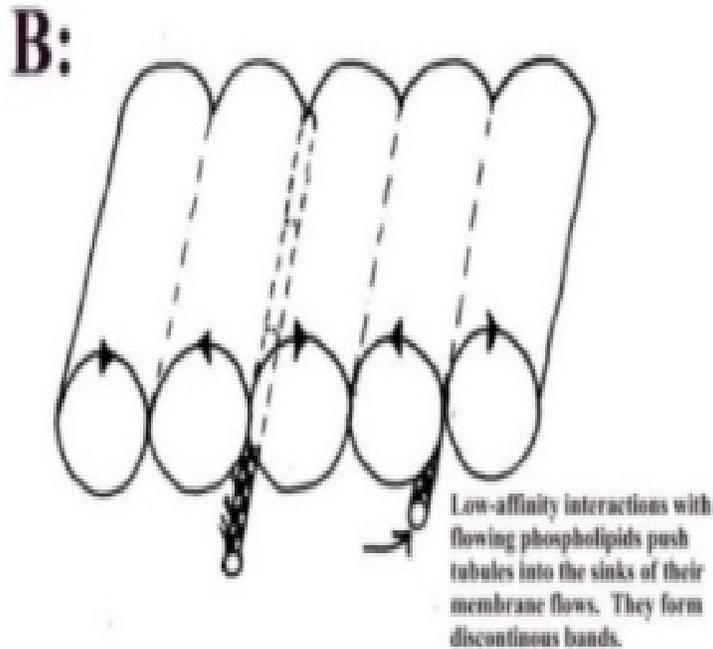

**B:** Any protein that is acylated, and has the capacity to undergo low-affinity interactions with phospholipids will be pushed into flow 'sinks' when it arrives at the membrane surface. As the plant cell is assembled therefore, Microtubules will form into bands, at a spacing that is equal to phospholipid flow wavelength. Providing the cell remains metabolically active, Microtubules will maintain this spacing and orientation relative to the cell axis.

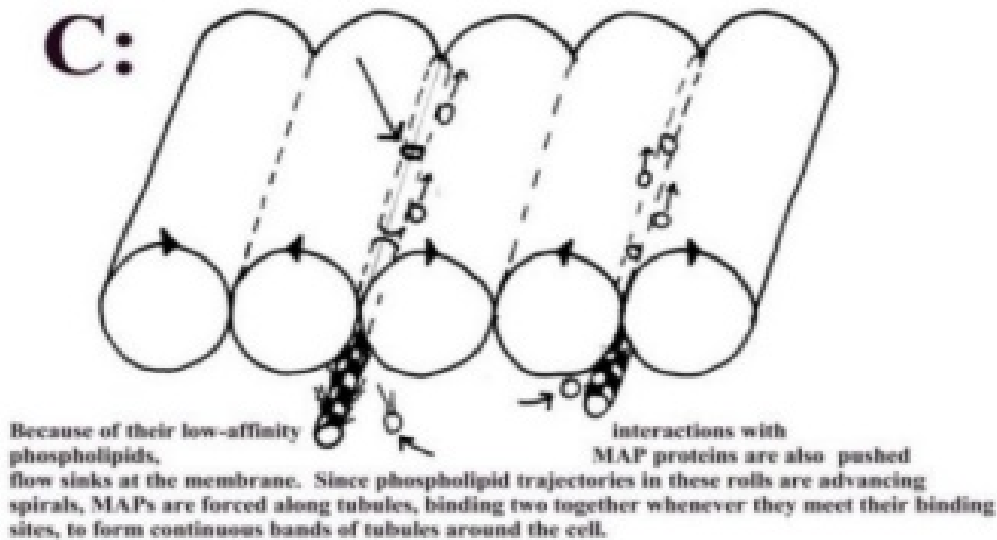

C: 'Microtubule Associated Proteins' are also acylated, and undergo low affinity interactions with phospholipids. These too will be pushed into flow sinks at the inner membrane surface. The advancing spiral trajectories of phospholipid molecules will exert a force upon these, and as they are pushed along an individual tubule band, they will bind to them whenever they encounter a binding site. In this manner, continuous Microtubule bands can be bound together.

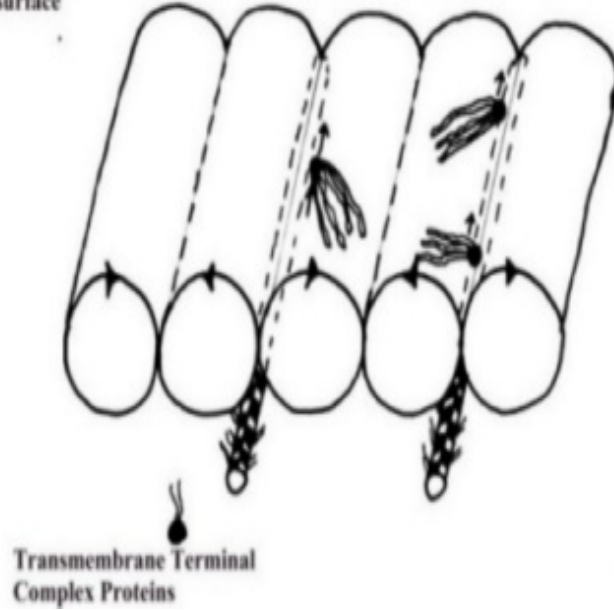

**D:** The next stage of plant cell assembly involves the delivery of Cellulose synthesising 'Terminal Complexes' to the membrane surface in lipid vesicles. According to the feedback model, these fuse with the membrane, and alter its' lipid composition. Since the proteins they contain bind phospholipid with low affinity, they too will be pushed into flow sinks at the inner membrane surface, and will also be forced along individual microtubules. Since some of these elements run through the membrane, the short Cellulose microfilaments that they produce as they progress are extruded onto the membrane surface. Regardless of where these Cellulose fibrils (which also bind phospholipids with low affinity) land on the cell surface, they will be pushed into flow sinks at *the outer membrane surface*, ie exactly mid-way between the inner cortical Microtubules (**E**) .

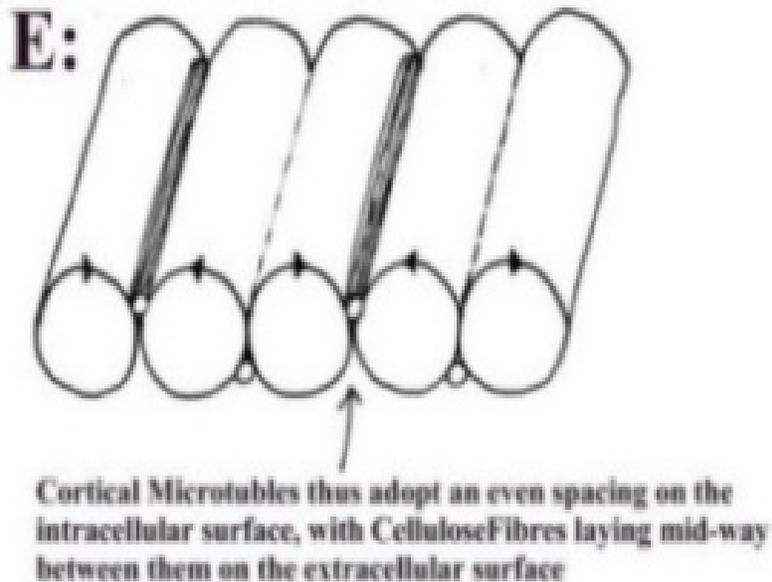

**E:** The model presented here may also explain why Cellulose produced *In Vitro* does not adopt its' native structure ([121](#) see [http://www.botany.utexas.edu/facstaff/facpages/mbrown/Phildoc/default.htm](http://www.botany.utexas.edu/facstaff/facpages/mbrown/Phildoc/default.htm).). *In Vitro*, (l~3)-fl-glucans (Callose) is synthesized instead of (l~4)- - glucans (Cellulose) when there is any disturbance of the plasma membrane. The model suggests this is because membrane lipid flows play a crucial role in determining the structure of the fibre and hence in determining its' tensile properties. Just as with erythroid Spectrin, where organised phospholipid flows determine the extent to which individual subunits of the cytoskeletal network edges are crumpled together, phospholipid microflow streamlines in individual rolls will determine the lateral alignment of Cellulose fibrils in the final fibres (Fig. 4F):

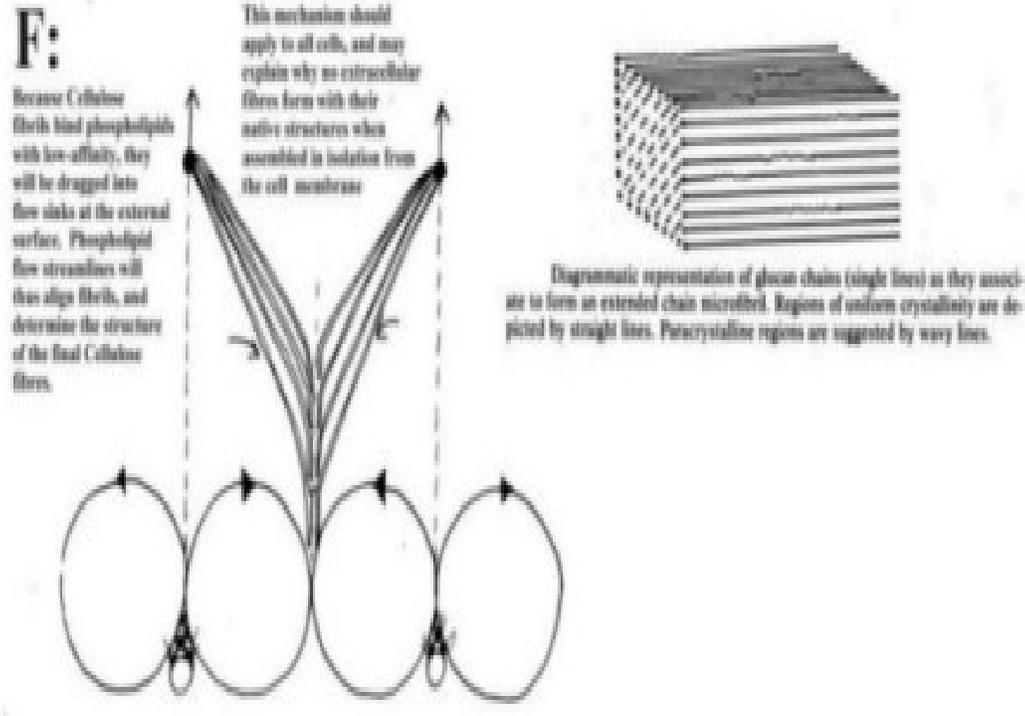

In support of this view, a recent publication by Brown et al. demonstrates that when a cell is half immersed in a viscous, inert fluid, it produces helical Cellulose on the immersed side, and Cellulose with normal structure on the exposed side. Carpita [158] has produced (l~4)- -glucans from detached cotton fibers but only by making adjustments to the viscosity of the extracellular medium with chemically inert polyethylene glycol.

Numerous other extracellular cytoskeletal components are also arranged parallel to internal features *In Vivo*, but also fail to adopt their native structure when they are produced *In Vitro* ([119, 121]): Tropocollagen for example, forms parallel or antiparallel arrays *in vitro* instead of the quarter-staggered arrays it forms *In Vivo* (for details see [123,124,125,161.] As all these fibres undergo low-affinity interactions with phospholipids, a role for fluid microflow streamlines as a template in their assembly *In Vivo* seems a viable hypothesis.

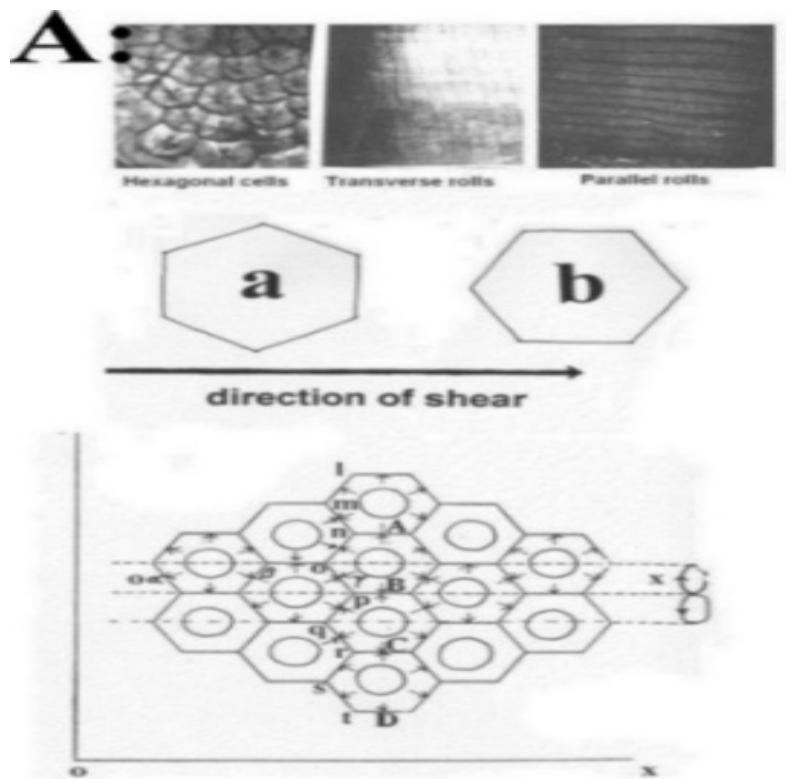

Fig. 3. Effects of shear on convecting viscous media (after Graham, 43)

A. Graham 43 demonstrated that application of surface shear can cause a bifurcation in stable polygonal convection system, and that shear rate determines whether the induced roll pattern run transverse or parallel to the direction of applied shear. Importantly, the *orientation* of the initial hexagonal cells with respect to applied shear force determined whether longitudinal rolls formed: hexagons meeting the direction of shear in orientation (**a**) deformed, those in orientation (**b**) did not at the shear rates studied. Likewise, the application of shear to a convecting system manifesting stable roll patterns can cause them to re-orient by 90 º (passing through a square cell intermediate pattern), depending on orientation

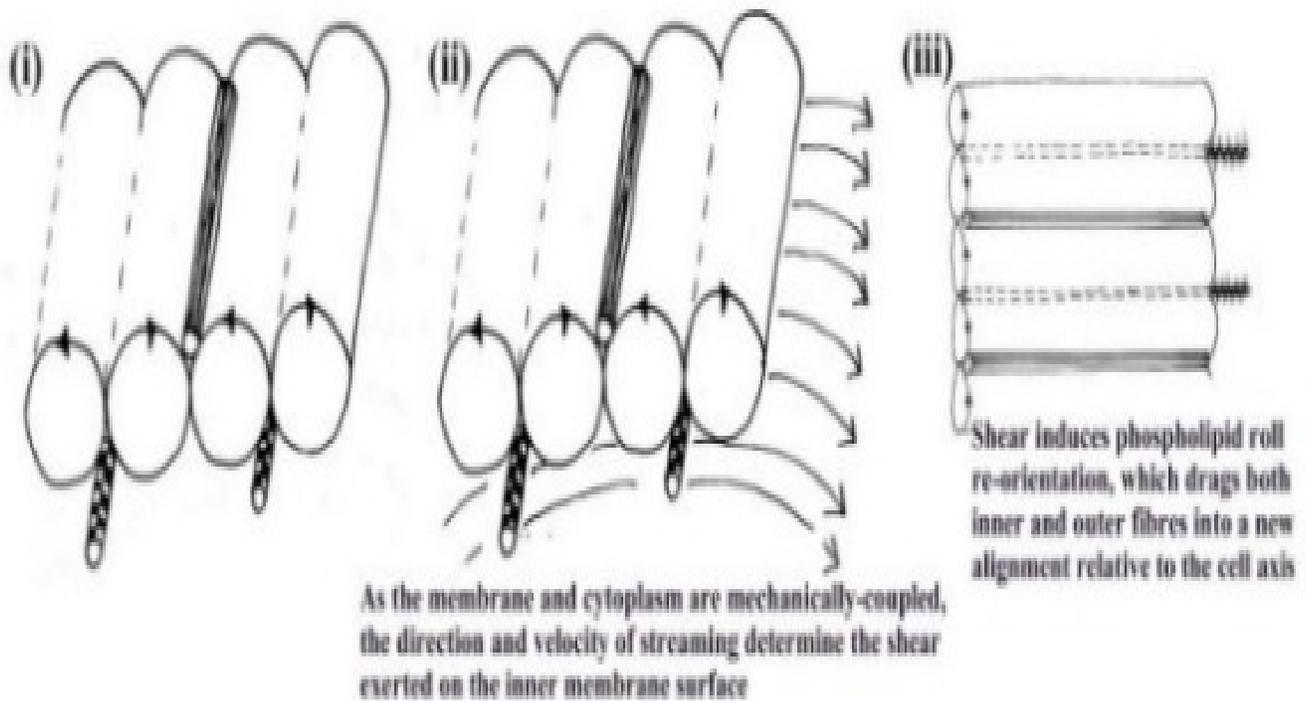

## B: Shear-induced re-orientation of Microtubules and Cellulose microfibrils.

Since organised phospholipid flows through the membrane (**i**) are mechanically coupled to the cytoplasm that flows beneath it, the latter could similarly exert shear (**ii**), causing stable convective phospholipid rolls to re-align . This would 'drag' cortical protein elements into new alignments (shown here transverse to parallel), determined primarily by the direction and velocity of cytoplasmic flow (**iii**). It is accepted that both the direction and velocity of cytoplasmic streaming varies from cell to cell, at specific stages in the lifespan of cells, and in response to changes in temperature 126, the gravity vector, and the viscosity of the suspending medium. For this reason I have suggested such flows are evidence of buoyancy-driven convection (see **Fig. 3C** below).

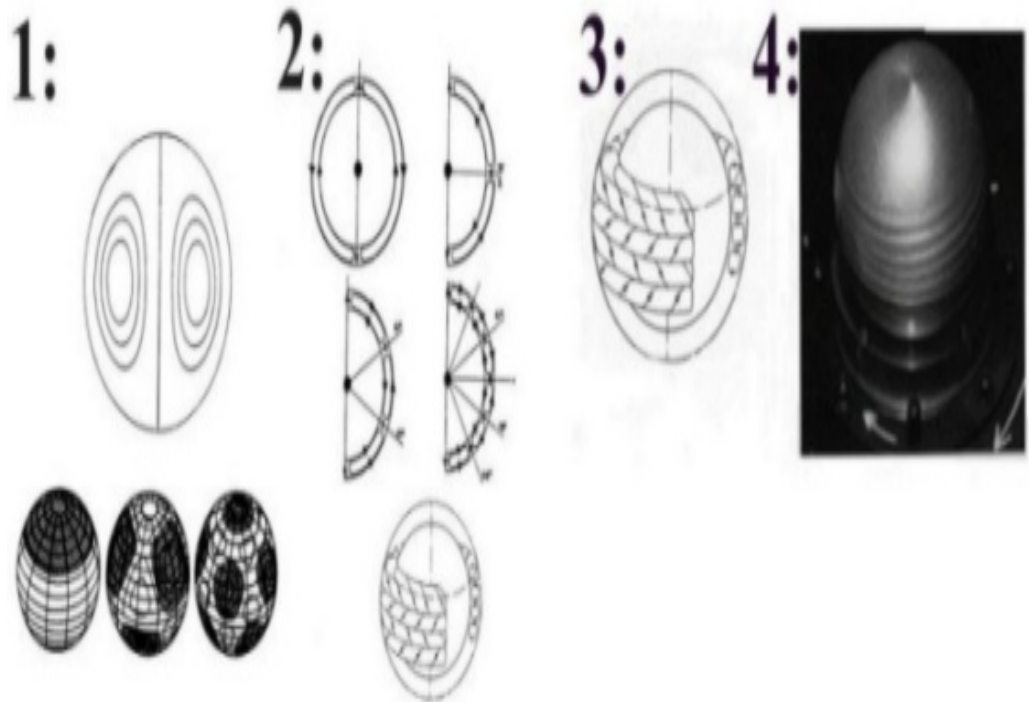

## C: Spatial Organisation in and on Spherical Fluid Domains

1. Four stable solutions to the Spherical Benard problem (After 90, 91, 92,127,128. When a sphere of viscous fluid is heated centrally, patterns of fluid up- (light) and down-wellings (dark) across it vary with radius: stable solutions can be axi-or non-axisymmetric.
2. Spherical shells of viscous fluids either heated from within 85,86,87 or

rotating (**3** and **4**: 88,89) also develop roll-patterned flows 156.

Biological cells can be envisaged as five immiscible layers of viscous fluid (nucleoplasm/nuclear membrane/cytoplasm/plasma membrane/fluid environment), the velocity and pattern in which convecting cytoplasm flows would thus determine the area(s) of membrane surface subjected to shear, and the force exerted.

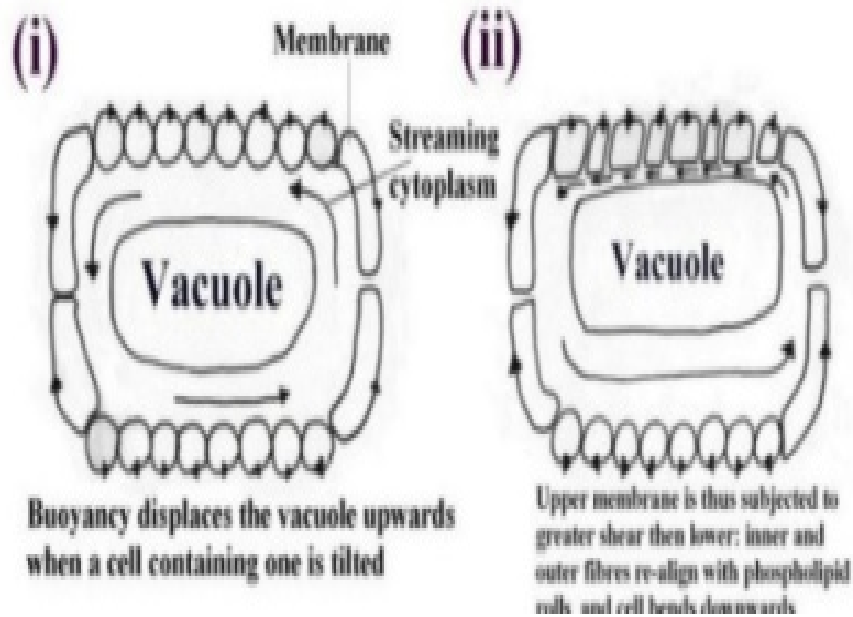

# Fig. 4: Shear-Induced Bifurcations as the Primary Cause of Gravitropic Bending

**A:** Gravitropic bending could be induced in cells with a vacuole by virtue of its' upward displacement, which would increase the shear experienced by the upper membrane surface Note: not all cell vacuoles are centrally situated.

**B:** Gravitropic bending is also a growth process: under this model, it could arise merely as a consequence of the cells' fluid dynamics. As the cell grows, its radius ratio increases and the composition (hence viscosity) of both immiscible fluid layers is altered. Extrapolating qualitatively from both the Spherical Benard and Spherical Couette problems (see Fig. 4C), the first parameter affects the stable pattern of up-and down-welling flows of fluid across the domain, the latter alters the extent of mechanical coupling between two layers. In a cellular context therefore, the areas of the membrane subjected to shear, its' direction and magnitude, will alter in the course of the cells' Developmental trajectory. The direction in which cells bend is determined by the area of membrane subjected to shear: as stable solutions need not be axisymmetric [91], *this mechanism allows all four geotropic responses* – positive, negative, transversal and plagiotropic – to be induced, in stark contrast to all previous hypotheses.

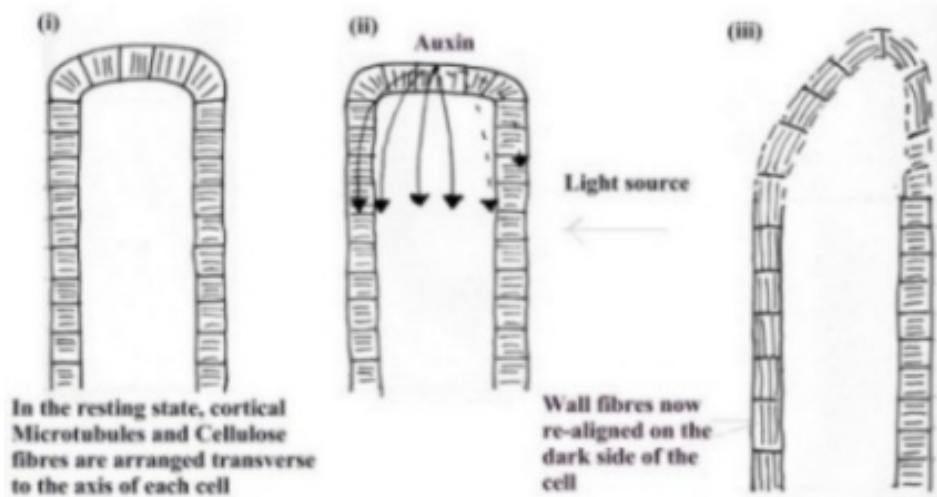

**C:** Phototropic bending is also a growth process that is accompanied by a 'transverse to longitudinal' re-alignment of the plant cells' Microtubules. It is known to involve Auxins (produced at the tip) being preferentially destroyed on the side exposed to a light source, but prior to this model, this morphogen has been assumed to exert its' effect exclusively by invoking conformational changes in cytoskeletal proteins. Under the new model, by altering the viscosity of the membrane layer in cells on the shielded side of the plant (see text), Auxins alter the extent of mechanical coupling between their cytoplasm and membranes, adjusting the amount of shear exerted, and causing re-orientation of phospholipid rolls (and hence of their wall fibres) such that unexposed cells undergo preferential elongation and the cells bend as shown.

In support of this model is the observation that treatment of cells with Auxins or UV light simultaneously alters the gravity- dependent polarity of cytoplasmic streaming 129,130: the new model is compatible with these observations, and suggests that both perturbations affects cytoplasmic streaming patterns *indirectly*, by exerting a primary effect on the lipid membrane, altering the amount of mechanical coupling between it and the cytoplasm in specific areas of the membrane, altering the amount of 'damping' that the flowing cytoplasm experiences at the same time as it changes cell shape.

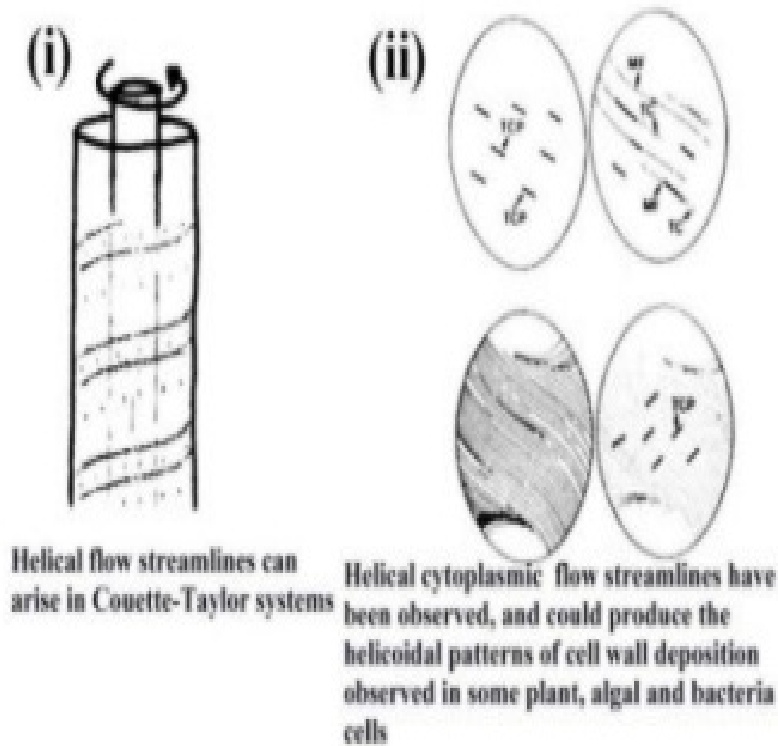

Fig. 5: Helical Cell Wall Deposition

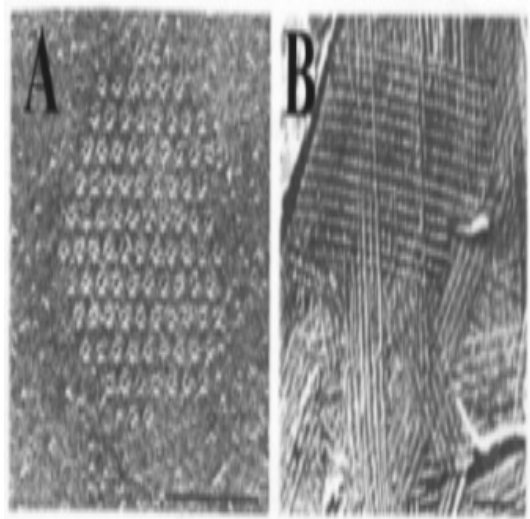

# Fig. 6: Arrangements of Terminal Complexes

Freeze fracture electron micrographs of *Micrasterias denticulata* semicells engaged in wall deposition.

**A:** Rows of TC rosettes in a hexagonal array show a constant centre-to-centre spacing of 28 nm, about twice the depth of the membrane.

**B:** Bands of parallel cellulose fibrils in the cell wall: the centre-to-centre spacing of fibres is also constant at 28 nm. From 8 with kind permission.

Under the new model, these two arrangements, seen in the same cell could evidence a shear-induced 'hexagon to roll' bifurcation in phospholipid flows through the cell membrane takes place during Development (see Appendix III).

---

# References

1. Lofthouse, J.T.(1998a) "The Dynamic Template Membrane"

2. Lofthouse, J.T. (1998b): Pattern Formation In Biological Fluids I : How Protein-Lipid Feedback Enables Membrane Flow Topology To Determine Biological Cell Shape

3. Lofthouse, J.T. (1998c): Pattern Formation In Biological Fluids II: How Bifurcations in Membrane Lipid Flow Planiforms Induce Cell Shape Changes In Shear Flows

4. Lofthouse, J.T. (1998d): Pattern Formation In Biological Fluids III: Red Cell Echinocytosis : Turbulent Membrane Flows At The " Edge of Chaos"

5. Yuan, M., Shaw, P., Warn, R. and Lloyd, C. (1994). Dynamic re-orientation of cortical Microtubules, from transverse to longitudinal, in living plant cells. Proc. Natl. Acad.Sci. USA 91 pp 6050-6053.

6. Giddings, T.H. and Staehlin, L.A. (1982). Membrane-mediated control of cell wall microfibrillar order. In 'Developmental Order: Its' Origins and Regulation' (Eds Subrwelny, S., and Green, P.B.) pp 133-147. Alan R. Liss, New York.

____________________________________________________________________

# Appendices:

**Appendix I:** The Dynamic Template model predicts that only uni-directional flow can be generated in a particular region of the membrane in any one layer. This in turn demands that all TCs move in the same direction across the cell until phospholipid rolls re-orient wrt the cell axis. It is only at odds with *one* published micrograph, 122, which shows that some of the 'paired' microfilaments in *Oocystis* have TCs on opposite ends, and concludes therefore that fibrils must be deposited in both directions simultaneously. As stated earlier, because electron microscopy produces only one 'freeze frame' image, and the same micrograph clearly shows single fibrils, and isolated TC's that are not attached to fibrils, I dispute the interpretation of this data. Whilst some of the paired fibrils appear to have TC at opposite ends , others in the same micrograph do not. The centre to centre spacing of paired fibrils is ~ 23.5 nm: under the new model, this would be equivalent to the estimated wavelength of phospholipid rolls (circa twice the depth of the membrane) suggesting that each fibril is being produced by a TC trapped between separate roll sources. The new model allows fibres to be laid down in separate directions ie TCs can move away from one another towards both poles, but this demands that cytoplasmic flows define at least a 2-torus, flowing outward at the cells' equator and in again at the respective poles. Sadly, there is insufficient high resolution data to confirm this.

**Appendix II:** In previous papers 1-4, I have suggested that there is sufficient empirical evidence to support the notion that cytoplasmic flows are evidence of buoyancy-driven convection. Because this phenomenon dependent on the gravity vector 134, exposure to a microgravitational enviroment will affect flow dynamics. Both the direction and velocity of cytoplasmic streaming are noticeably affected when plant cells are suspended in a medium of higher viscosity, and when exposed to a microgravitational environment (reviewed in 75. The former effect clearly implicates mechanical coupling of the intracellular cytoplasmic fluid to the outer environment, mediated through the lipid membrane. The latter is compatible with behaviour predicted in buoyancy-driven flows. Whilst there is no sound equation to calculate the velocity of convecting fluid 134, in the linear approximation, flow velocity is proportional to e$\omega^t$, where t is the time, and $\omega$ is determined by the Rayleigh number. Hence, from Linear Theory, since velocity is inversely proportional to G, the rate of cytoplasmic streaming should increase on exposure to a microgravitational environment - both acropetal up and basipetal streaming velocities are observed to *increase* on exposure to 10$^{-4}$ G 152,75, supporting the notion of buoyancy-driven flow.

Convective processes have previously been excluded as pattern-forming mechanisms in biological systems for several reasons:

**(i) The 'Linear Theory' Argument.**

Linear Theories apply only to 'Natural' or 'Free' convection in thin horizontal layers of infinite extent heated uniformly from below, where flow is entirely a response to forces acting in the fluid. In a living cell, cytoplasm and lipids are mechanically/thermally coupled, and flow past the corrugated surfaces of proteins: since these are predominantly helical, their surfaces will induce deviations from their 'natural' planiforms. ATP and GTP hydrolysis at specific points on the cytoskeletal lattice will also induce localised 'heating', another source of forcing. Linear theories therefore cannot meaningfully be used to calculate the temperature gradient that a cell would need to establish across its' radius for convective onset to arise.

Where this has been done in the past, using the approximate viscosity of lipids, the depth of biological membranes ($\cong$ 5-18 nm for the lipid membrane) and typical cell radii (1-20 μm), they indicate that unfeasibly high, (non-physiologically realisable) temperature gradients ($\cong$ 100 °C) would have to be established across the cell for the onset of either buoyancy- or surface tension- driven convection. For this reason, I focus on presenting a good *qualitative* model, rather than a poor quantitative one. As currently formulated, Linear Theory fails to explain various phenomena even in 'Natural' convection, for example, the empirical effect of temperature on convective wavelength. In fluids composed of more than one molecular species (binary fluids), convective onset is characteristically *sub-critical* 135, and in nematics, appears at a threshold temperature up to $10^3$ times smaller than theory predicts for an isotropic fluid of comparable average properties, because flow alignment reduces thermal conductivity 136. Silicon oils are multicomponent systems of more than one type of molecule, and even these exhibit 'Natural' convective behavour that cannot be predicted by current theories ( see 138. Since with lipids, the headgroup type and acyl chain length/structure vary from cell to cell in biological cells, and undergo constant remodelling during lifespan, it is impossible to use a value of the kinematic viscosity, and calculate the temperature gradient needed for onset as others appear to have done in the past (eg Tabony and Job 139). In the absence of any strong data to suggest otherwise, I suggest that cytoplasmic streaming could be organised by Rayleigh-Benard convection, and that this, via feedback with protein and the membrane, provides the cell with a means of transducing the gravity vector into biological form.

One publication has proposed that biological systems show gravity dependence by way of the bifurcation properties of certain types of non-linear chemical reactions ('Reaction-Diffusion' RD mechanisms: Turing,102; Prigogine and Stengers 140)and appears to suggest that it is impossible for bouyancy driven convection to arise in a cell. Work presented in Lofthouse 1 demonstrated that RD is insufficient to account for the pattern of the chief red cell membrane protein Spectrin, and here is shown to be unable to account for pattern formation in plant cells. Simply stated, typical protein diffusion constants yield patterns of fixed wavelength on the micron scale: whilst this may appear convenient when addressing pattern scales at a tissue level, cytoskeletal banding patterns in plant cells **are three orders of magnitude lower** than this. Turing himself acknowledged that RD might be supplemented by other mechanisms *In Vivo*. It is important therefore to critically dissect a set of publication that claims to have demonstrated that buoyancy-driven convection is not involved Microtubule pattern formation. Tabony and Job 139 and Papaseit 141suggested that In Vitro, aggregates of Tubulin protein (that they refer to as 'Microtubules') show patterns of orientation that are gravity-dependent *In Vitro*, and moreover, suggest their observations can be extrapolated to pattern formation In Vivo. This work must be criticised from both a Fluid Dynamics and a Biology standpoint.

This work has tenuous physiological relevance: the tubulin structures these workers describe as 'Microtubules' are visible to the naked eye, a staggering 1 mm thick (compared to the ~25 nm diameter microtubule cylinders formed in living, metabolically active cells). These are **aggregates of Tubulin**: those formed within the first 15 minutes of their *In Vitro* assay show *no preferred* orientation: strong orientational ordering only developed slowly over ~ 5 hours, as these aggregates increased in size. Tabony 139attempts to show that buoyancy-driven convection is not involved exclusively by using Linear Theory to calculate the temperature gradient across their reaction vessel, assuming rigid-rigid boundary condition. This only applies to fluids of infinite horizontal extent heated from below: once tubulin aggregates have started to form, they define the spatial set on which heat is dissipated, and their system is no longer 'heated from below' but are heated **from within.-** the limitations of extant theory are explained in Velarde & Normande (1980). In laboratory situation, both the onset of convection and the pattern selected by the fluid are dramatically affected by boundary conditions and the systems' '**Aspect Ratio'** (essentially the ratio of fluid depth to horizontal extent : 24, 25). This factor is not considered in their paper. Their reaction vessel is a spectrophotometer cell (4cm x 1cm x 1mm): when standing up, d = 4cm, when laying down horizontally and subject to rotation, d = 1mm, consequently, the aspect ratio of their experiment differs in the control.

An additional problem in extrapolating their observations to *In Vivo* assembly arises from their use of $^2H_2O$ buffers. They state this has no effect on Microtubulin reaction dynamics, however, *In Vitro*:it is known to prevent the depolymerisation of Microtubules142, to prevent the depolymerisation of microtubules Colchicine 143, to cause a demonstrable increase in the number of Microtubules formed in Sphagnum leaflets 17, to interrupt growth and causes irregular wall thickening in Funararia caulonema cells 144, to delay both the onset of germination and interestingly, to alter the *symmetry* of germ tubes formed in spores . Buoyancy-driven convection as a G-sensitive pattern forming mechanism *In Vivo* cannot therefore be excluded by Job and Tabonys' data. Whilst Reaction-Diffusion schemes can lead to the formation of stationary patterns that resemble the Microtubule lattice, this approach have so far failed to explain the finer details of Microtubule *structure* or the dynamics of their assembly. Because the assembly of multi-subunit proteins (eg Actin, Tubulin) requires ATP or GTP hydrolysis, it is exothermic, and a substantial amount of energy is released into the embedding fluid as it progresses. Injecting heat into viscous fluids causes convective flows, and in certain circumstances, these forms spatially defined structures. Moreover, these flows patterns are affected by the contours of the surfaces that the fluid flows past 145. The feedback mechanism presented in this series of papers demonstrates how pattern formation in the cells fluids plays a crucial role in assembling their proteins. Similar feedback considerations allow one to extend this mechanism to pattern formation in the cytoplasm. Just as the polygonal geometry of the erythroid Spectrin lattice reflects the convective aminophospholipid structures against which it forms, structural details of the intermediates formed during tubulin assembly also appear to mimic classic fluid structures, and this, coupled to the kinetics of the assembly process, suggest that Microtubules are actually formed *against the surface* of dynamic, dissipative fluid vortices. This is detailed in Lofthouse1,2,3,4.

**(ii) Nicholis and Prigogine's 146 incorrect assumption that biological systems are "isothermal", and subject to "constant, time-independent boundary conditions":**

Far from being 'isothermal', fluid regions within a cell volume in which ATP/GTP are being hydrolysed are 'hotter' than regions that are not**:** thermal *gradients* therefore exist across both the cytoplasm and membranes: sub-critical onset could arise with a gradient as low as 0.1 ºC. As detailed above, cell boundary conditions are not 'constant': lipid composition (and hence viscosity) change continuously with time, and cells at the centre of a mass of tissue have vastly different boundary environments than those in its' periphery. With the membrane viewed as nothing more than a random, hydrophobic fluid, these factors appear insignificant, but under the feedback model presented in this paper, acyl chain composition determines membrane viscosity, and phospholipid content determines the number and strength of dipole-dipole interactions: lipid composition therefore defines both the critical Rayleigh- *and* Marangoni numbers of the membrane, and hence the point of onset for organised flows. Neither are the boundary conditions constant: red cells begin development in a hotter viscous environment than the plasma, where they are subjected to constant shear. Since the cytoplasm, membrane and environment are clearly capable of mechanical and/or thermal coupling, a flowing cytoplasm can exert shear on the membrane, a factor that can have a dramatic effect on flow planiform

Cytoplasmic streaming direction and velocity are both sensitive to the gravity vector, temperature, ATP concentration (ie the amount of metabolic heat dissipation), the viscosity of the extracellular and intracellular media , as are tropic responses [106] – all factors that would affect buoyancy-driven convection which I suggest is the phenomenon being observed. Streaming is discerned by observing the motion of intracellular solids using light microscopy. Depending on cell type, stage of Development, and environmental factors (gravity, temperature and intriguingly, the viscosity of the fluid in which the cell is suspended), these objects (starch granules, nuclei, vesicles) move upwards, downwards and around in the cell in defined rather than random trajectories. As interpreted by Cell Biologists using their Statics-focussed cell models, because cellular fluids are assumed 'random', and (according to their facile understanding of it)'Gravity can only act downwards' (!), the only way an object could move against the gravity vector in a definite path is to be physically pulled along a protein 'guide rail' by unseen 'contractile' protein elements. I stress that such 'winch and pulley' mechanisms are only *assumptions*, formed in ignorance of the existence of spatially ordered fluid flows. If we were to observe the bending motion of a tree through a window, without including the presence of the viscous fluid in which it is embedded in our model, we explain its movements entirely in terms of the observable 'Statics', and would invent a contractile tree apparatus. All previous models of gravi- and photo-tropic bending are this facile. If we know the tree is surrounded by a viscous medium flowing with a defined trajectory, however, we conclude that the tree is not contractile, but is being 'folded' by the spatially organised fluid flows in which it is embedded. The viscous fluids (membrane/cytoplasm/nucleoplasm) in which cells statics (protein/DNA) are embedded are never incorporated in models of cytoplasmic streaming: it is the assumption that their motion is 'random' that ultimately produces a 'Statics' based model of moving objects in the cell.

Such interpretations of streaming phenomena suffer from several flaws:-

(i) Microtubules and/or Actin filaments are invariably suggested as the structures that cellular inclusions move along like 'ratchets and pinion'. The assumption that Colchicine only affects proteins are largely to blame for this mis-interpretation. When added to cells, this reagent inhibits cytoplasmic streaming ie is observed to alter the motion of objects in intact cells Colchicine is suggested to inhibit the motion of objects in the cell by disrupting the Microtubules that cellular inclusions 'walk' along. The fact is that Colchicine does not bind to intact Microtubules. An invariably omitted fact is that it also binds phospholipids, altering their phase transition temperatures. That it causes mature human red blood cells to change shape, despite the fact that these have no Microtubules whatsoever indicates that it is exerting its' effect in some way other than by dissembling Microtubule 'guide rails'. Under the new feedback model, Colchicine alters convective organisation of aminophospholipids: this alters the mechanical coupling between the cytoplasm and membrane, and thus removes the 'brake' on cytoplasmic flows. Colchicine is known to inhibit the shape changing effect of Gibberellic acid and the gravitropic responses of many plant cells, whilst treatment with Auxins eliminates the gravity-dependent polarity of cytoplasmic streaming ( [163, 164]) - I suggest this is because of its' affect on the dynamics of cellular fluids, rather than its' binding to specific proteins. Treatment of cell membranes with anaesthetics (which bind phospholipids and alter their phase transition temperatures: [149] also inhibits cytoplasmic streaming: under my new model, this would appear to arise by them altering the extent of mechanical coupling between the cytoplasm and the membrane.

(ii) The second problem with 'winch and pulley' interpretations of cytoplasmic streaming phenomena is that it presents a recursive problem. If an object can only be moved in a cell from point A to B along a strand of Actin or a Microtubule running between those points, the cell requires a pattern-forming mechanism competent to place protein in that spatial position: no viable mechanism has ever been posited.

Cytoplasmic streaming is a dynamic phenomena: it is senseless to expect to understand it by observing only the cells' Statics. It is alterations in the cells' dynamic parameters that affect its' rate and direction. Streaming is only observed if the cell is metabolically active, ie generating heat internally. Because Rayleigh-Benard convection in a spherical domain can produce acropetal-, basipetal-, upwelling- and downwelling-flows of fluid that are either axi- or non-axisymmetric (Fig. 3C), I suggest that the observed motile inclusions could actually be transported in the directed, spatially-organised convection streams it generates. The upward and downward velocities of moving objects in the cell need not be equal, since streamlines with differing velocities are present in each convection cell, hence the velocity of a moving object would be determined by the fluid streamline it was trapped in.

**Appendix III:** The gametes of algae and lower plants usually lack a Cellulose cell wall: whilst some have scales, spines or a cucilaginous layer, these have no apparent influence on shape. Bouck and Brown,[147] showed that the application of hydrostatic pressure causes these cells to become spherical: removal of the pressure leads to a restoration of cell shape. Application of external pressure is known to depolymerise microtubules formed *in vivo* [148] but also affects the phase transition temperature of PC membranes ([166]. Under the Dynamic Template model, the application of pressure may be seen as a change in the boundary of the cell, which according to Linear theory would affect the onset of convection, the spatial distribution of MF, and hence cell shape, and is thus entirely compatible with the feedback mechanism of cell shape I have suggested.